\documentclass[twocolumn,showpacs,amssymb,amsmath,aps,floatfix,prl,nofootinbib,superscriptaddress]{revtex4-1}

\usepackage{hyperref}
\usepackage[all]{hypcap}
\usepackage{array}
\usepackage{graphicx}
\usepackage{amsmath,amsthm,amssymb}

\usepackage{color}

\begin{document}

\title{Gradient Tomography of Jet Quenching in Heavy-Ion Collisions}

\author{Yayun He}

\affiliation{Key Laboratory of Quark \& Lepton Physics (MOE) and Institute of Particle Physics, Central China Normal University, Wuhan 430079, China}

\author{Long-Gang Pang}
%\email{lgpang@lbl.gov}
\affiliation{Key Laboratory of Quark \& Lepton Physics (MOE) and Institute of Particle Physics, Central China Normal University, Wuhan 430079, China}

\author{Xin-Nian Wang}
\email{xnwang@lbl.gov}
\affiliation{Key Laboratory of Quark \& Lepton Physics (MOE) and Institute of Particle Physics, Central China Normal University, Wuhan 430079, China}
\affiliation{Nuclear Science Division, Lawrence Berkeley National Laboratory, Berkeley, CA 94720, USA}
%\affiliation{Physics Department, University of California, Berkeley, CA 94720, USA}

\begin{abstract}
Transverse momentum broadening and energy loss of a propagating parton are dictated by the space-time profile of the jet transport coefficient $\hat q$  in a dense QCD medium. The spatial gradient of $\hat q$ perpendicular to the propagation direction can lead to a drift and asymmetry in parton transverse momentum distribution. Such an asymmetry depends on both the spatial position along the transverse gradient and path length of a propagating parton as shown by numerical solutions of the Boltzmann transport in the simplified form of a drift-diffusion equation. In high-energy heavy-ion collisions, this asymmetry with respect to a plane defined by the beam and trigger particle (photon, hadron or jet) with a given orientation relative to the event plane is shown to be closely related to the transverse position of the initial jet production in full event-by-event simulations within the linear Boltzmann transport model. Such a gradient tomography can be used to localize the initial jet production position for more detailed study of jet quenching and properties of the quark-gluon plasma along a given propagation path in heavy-ion collisions.
\end{abstract}



\maketitle

\noindent{\it 1. Introduction}: --
Parton energy loss inside the quark-gluon plasma (QGP), a deconfined form of nuclear matter created in high-energy heavy-ion collisions, is predicted to cause the suppression of high transverse momentum jet and hadron spectra, also known as jet quenching \cite{Gyulassy:1990ye,Wang:1991xy}. The discovery of jet quenching at the Relativistic Heavy-ion Collider (RHIC) \cite{Adcox:2001jp,Adler:2002xw,Adler:2002tq} and the continued study at the Large Hadron Collider (LHC) \cite{Aad:2010bu,Chatrchyan:2012gt,Chatrchyan:2013kwa,Aad:2014bxa} provide unprecedented opportunities to explore detailed properties of QGP. One of these properties that are related to jet quenching is the jet transport coefficient $\hat q$ \cite{Baier:1996sk,CasalderreySolana:2007sw} or the transverse momentum broadening squared per unit distance due to jet-medium interaction. The extraction of the jet transport coefficient from the combined data on the suppression of single inclusive hadron spectra at both RHIC and LHC by the JET Collaboration \cite{Burke:2013yra} and similar efforts \cite{Andres:2016iys,Chen:2016vem,Cao:2018ews,Feal:2019xfl} is a clear demonstration of the power of jet quenching for the study of QGP properties.

In the current extraction of the jet transport coefficient, one assumes the spatial distribution of the initial jet production position according to the Glauber model with the Woods-Saxon nuclear distribution and the space-time evolution of the bulk QGP matter according to a relativistic hydrodynamic model. The azimuthal anisotropy of jet quenching \cite{Wang:2000fq,Gyulassy:2000gk} can shed light on the averaged path-length dependence of jet transport. However, the study of jet quenching with a localized initial jet production position can provide more direct information about the space-time profile of the jet transport coefficient.  Although single hadron, dihadron and $\gamma$-hadron measurements allow one to select events with surface, tangential and volume emission of large transverse momentum hadrons \cite{Renk:2006qg,Zhang:2007ja,Zhang:2009rn}, one still cannot localize the initial jet production position in the transverse plane of the heavy-ion collisions. In this Letter, we propose and demonstrate a gradient tomography with which one can approximately localize the initial jet production position in the transverse plane. 

Since the jet transport coefficient $\hat q$ characterizes the momentum broadening transverse to the jet propagation direction, it determines the diffusion in both the transverse momentum and coordinate of a propagating jet parton. If the spatial distribution of $\hat q$ is not uniform, its transverse gradient will lead to a drift in the final parton's transverse momentum and spatial distribution. One can define a momentum asymmetry in the transverse direction to characterize this drift which will depend on the propagation length and transverse gradient of $\hat q$.  Given the geometry of the bulk medium and its time evolution, one can in return use this transverse momentum asymmetry  to localize the initial production position of the selected jets in the transverse plane. We will define such a transverse momentum asymmetry and use a simplified Boltzmann transport equation and the full linear Boltzmann transport (LBT) model \cite{Li:2010ts,Wang:2013cia,He:2015pra,Luo:2018pto} to demonstrate the concept of a gradient tomography to localize the transverse position of initial jet production for a more detailed study of jet quenching.

\noindent{\it 2. Drift-diffusion equation}: --
The transport of a parton in a thermal QGP medium through elastic scattering can be described by the Boltzmann equation,
\begin{eqnarray}
k_a\cdot\partial f_a&=& \sum_{b c d } \prod_{i=b,c,d}\int\frac{d^3k_i}{2\omega_i(2\pi)^3} (f_cf_d-f_af_b) \nonumber \\
& \times & |{\cal M}_{ab\rightarrow cd}|^2\frac{\gamma_b}{2}(2\pi)^4\delta^4(k_a\!+\!k_b\!-\!k_c\!-\!k_d),
\label{bteq}
\end{eqnarray}
where $|{\cal M}_{ab\rightarrow cd}|^2$ are the leading-order (LO) elastic scattering amplitudes \cite{Eichten:1984eu},  $f_i=1/(e^{k_i\cdot u/T}\pm1)$ $(i=b,d)$ are phase-space distributions for thermal partons in QGP with local temperature $T$ and fluid velocity $u=(1, \vec{v})/\sqrt{1-\vec{v}^2}$, and $\gamma_b$ is the color-spin degeneracy for parton $b$. The summation is over all possible parton flavors and scattering channels. If the interaction is dominated by small angle scattering, one can expand the reaction rate in the momentum transfer of each scattering \cite{Prino:2016cni} and further neglect the effect of the flow and drag term due to elastic energy loss, the transport equation becomes,
\begin{equation}
\frac{k^\mu}{\omega}\partial_\mu f_a(\vec k,\vec r)=\frac{\hat q_a}{4}\vec\nabla_{k_\perp}^2 f_a(\vec k,\vec r),
\label{bteq2}
\end{equation}
where the jet transport coefficient $\hat q_a$ is defined as
\begin{eqnarray}
\hat q_a&=& \sum_{b c d } \prod_{i=b,c,d}\int\frac{d^3k_i}{2E_i(2\pi)^3} f_b(k_b) (\vec k_{a\perp}-\vec k_{c\perp})^2 \nonumber \\
& \times & |{\cal M}_{ab\rightarrow cd}|^2\frac{\gamma_b}{2}(2\pi)^4\delta^4(k_a\!+\!k_b\!-\!k_c\!-\!k_d),
\label{qhatdef}
\end{eqnarray}
or the average transverse momentum transfer squared per mean-free path. We can further assume the transverse momentum is small $k_\perp/\omega\ll 1$ and redefine the time along the light-cone $t\rightarrow t-z$, the above equation can be cast into a drift-diffusion equation in the transverse direction,
\begin{equation}
\frac{\partial f_a}{\partial t}+\frac{\vec k_\perp}{\omega}\cdot \frac{\partial f_a}{\partial \vec r_\perp}=\frac{\hat q_a}{4}\vec\nabla_{k_\perp}^2 f_a(\vec k,\vec r).
\label{eq:diff}
\end{equation}
A solution for the above drift-diffusion equations with an initial condition $f(\vec k,\vec r,t=0)=(2\pi)^2\delta^2(\vec{r}_\perp)\delta^2(\vec{k}_\perp)$ in a static and
uniform medium (where $\hat q_a$ is a constant) can be found with the method of Fourier transformation in both the transverse momentum and coordinate,
\begin{eqnarray}
f_a&=&3\left(\frac{4\omega}{\hat q_a t^2}\right)^2 
\exp\left[-(\vec{r}_\perp-\frac{\vec k_\perp}{2\omega}t)^2\frac{12\omega^2}{\hat q_a t^3}-\frac{k_\perp^2}{\hat q_a t}\right].
\label{eq:solution}
\end{eqnarray}
This simple analytic solution describes the transport of a single parton with diffusion in both the transverse momentum and coordinate space. The diffusion widths are $\sqrt{\langle k_\perp^2\rangle}=\sqrt{\hat q_a t}$ and  $\sqrt{\langle r_\perp^2\rangle}=t\sqrt{(\hat q_a t/3})/\omega$ (after integrating over $\vec r_\perp$ or $\vec k_\perp$), respectively. There is also a drift $\vec r_\perp=(\vec k_\perp/2\omega)t$ in the transverse coordinate for a given finite transverse momentum.

 \noindent{\it 3. Gradient-driven transverse momentum asymmetry}:--
In a dynamic and nonuniform medium,  the jet transport coefficient $\hat q$ varies in both space and time. One has to solve  the drift-diffusion equation numerically.  As one can intuitively expect,  the transverse momentum distribution will develop an asymmetry in the direction of the gradient for a given initial position and propagation direction. One can therefore use such an asymmetry in the final parton spectrum to localize the initial jet production and study jet quenching with a given geometry and propagation path.
  
  \begin{figure}
\centerline{\includegraphics[width=8.0cm]{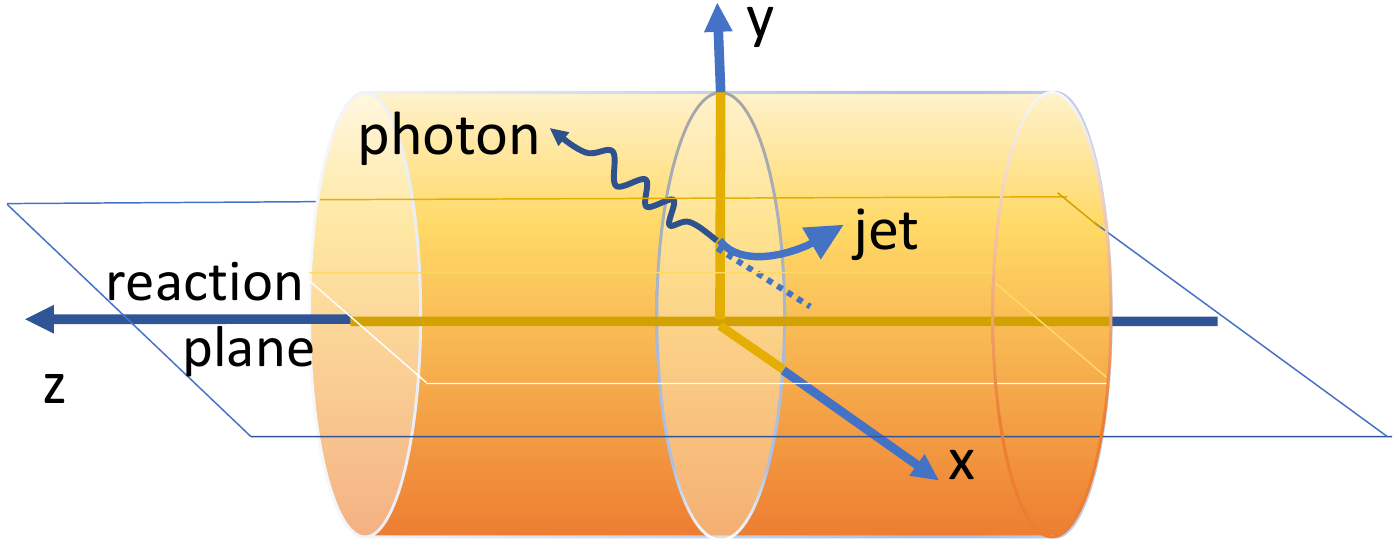}}
\vspace{-0.0 in}
 \caption{(Color online) Illustration of the transverse geometry and a  $\gamma$-triggered parton propagation in heavy-ion collisions.}
 \label{illustration}
\end{figure}

In high-energy heavy-ion collisions, the produced QGP matter has an initial transverse distribution given by the Glauber model of nuclear collisions. The longitudinal distribution has an approximate  Bjorken scaling with a uniform distribution in spatial rapidity $\eta=\ln[(t+z)/(t-z)]$. The transport of jet partons in the transverse plane through such a nonuniform matter will be influenced by the geometry of the QGP matter. 

Using the approximate back-to-back configuration of dijet, hadron-jet, or $\gamma$-jet events, one can select the direction of jet propagation relative to the reaction plane using the trigger particle as illustrated in Fig.~\ref{illustration}, where the trigger photon coincides with the reaction plane whose sign can be determined by $v_1$ of bulk hadrons at large rapidities. The local gradient of the jet transport coefficient will lead to a drift in the parton transport towards the less dense region of the matter. This drift will then result in a momentum asymmetry relative to the plane defined by the beam and the trigger particle (the sign of this plane $\vec n$ should be the same as that of the reaction plane),
\begin{equation}
A_N^{\vec n}=\frac{\int d^3rd^3kf_a(\vec k,\vec r) {\rm Sign}(\vec k\cdot \vec n)}{\int d^3rd^3kf_a(\vec k,\vec r)}.
\label{eq:asym}
\end{equation}
The sign and value of the drift and the momentum asymmetry will depend on the propagation length and the local gradient of the jet transport coefficient perpendicular to the propagation path which ultimately depend on the initial position and direction of the jet production. In return, with a given sign and value of the momentum asymmetry one can essentially localize the production position of the selected jet samples for a given propagation direction relative to the event plane. We refer to this localization of jet production using the transverse momentum asymmetry due to the gradient of the transport coefficient as the gradient tomography of jet quenching. 

\begin{figure}
\includegraphics[width=7.5cm]{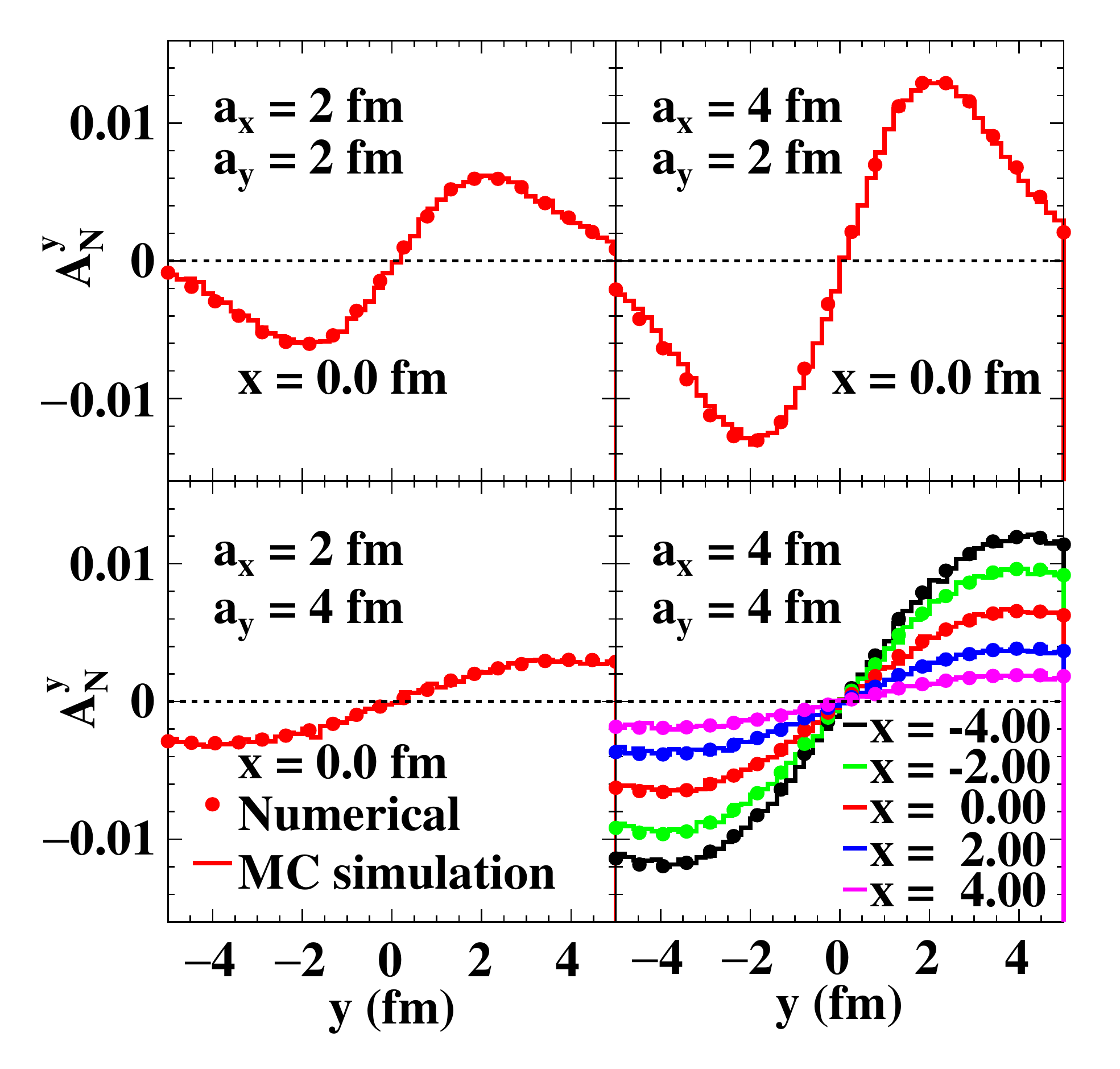}
\vspace{-0.0 in}
 \caption{ (Color online) Transverse asymmetry as a function of the initial transverse position $y$ for different position $x$ along the propagation path according to Eq.~(\ref{eq:diff}) with the jet transport coefficient in Eq.~(\ref{eq:simpleq}) and $\hat q_0=5$ GeV$^2$/fm. The histograms and solid circles are calculated from Monte Carlo and direct numerical method, respectively.}
 \label{simpleA}
\end{figure}

To illustrate this principle of gradient tomography, we numerically solve the drift-diffusion equation in Eq.~(\ref{eq:diff}) for a simplified space-time profile of the QGP matter in which the jet transport coefficient has a simple form of spatial and time dependence,
\begin{equation}
\hat q(\vec r_\perp,t)= \frac{\hat q_0 t_0}{t_0+t}e^{-x^2/a_x^2 - y^2/a_y^2}.
\label{eq:simpleq}
\end{equation}
We assume a parton with energy $\omega=20$ GeV is initially produced at $\vec r_{\perp}=(x,y)$ and propagates along the direction of the $x$ axis. Shown in Fig.~\ref{simpleA} are the transverse asymmetries of the final parton momentum distribution $A_N^y$ as a function of the initial position $y$ for different geometries $(a_x, a_y)$ and the initial position $x$ along the propagation direction. The initial value (at $t=0$ with $t_0=0.5$ fm/$c$) of the jet transport coefficient at the center of the medium is $\hat q_0=5$ GeV$^2$/fm. The asymmetry is seen to increase with the initial transverse position $y$ as the gradient increases until at the edge of the medium. The asymmetry is also bigger for in-plane propagation $(a_x,a_y)=(4,2)$ fm than for the out-plane propagation $(a_x,a_y)=(2,4)$ fm since the former has a larger gradient over a longer propagation length. The system size dependence of the asymmetry for central collisions $a_x=a_y$ is quite weak since a longer propagation length for a large system is offset by the smaller local gradient relative to a small system. The asymmetry also increases with the path length for fixed transverse position $y$ when the initial position $x$ along the propagation direction is varied. Since the asymmetry depends on both the transverse position $y$ and the propagation length or the position $x$ along its path, one should be able to use the asymmetry associated with a jet to localize its initial production position.

\noindent {\it 4. Linear Boltzmann transport model}: --
The energy and propagation length dependence of the asymmetry as illustrated by the solution of Eq.~(\ref{eq:diff}) will be modified by
the longitudinal diffusion in Eq.~(\ref{bteq2}). Inelastic scatterings and transverse flow will also further complicate the simple picture. We will use the LBT model \cite{Li:2010ts,Wang:2013cia,He:2015pra,Luo:2018pto} for a full simulation of the jet  transport in the QGP medium to further illustrate the concept of the gradient tomography in high-energy heavy-ion collisions. 

The LBT model was developed to study jet transport in the QGP medium in high-energy heavy-ion collisions. The model includes both $2\to 2$ elastic  parton scattering and inelastic processes of medium induced $2\rightarrow 2+n$ multiple gluon radiation for both jet shower and medium recoil partons according to the Boltzmann equation similar to Eq.~(\ref{bteq}). To account for the backreaction in LBT, the initial medium partons in each scattering are tracked as ``negative'' partons and propagate in the medium according to the Boltzmann equation. The energy and momentum of the negative partons will be subtracted from all final observables. These negative partons  and the medium recoil  partons are referred to as jet-induced medium response \cite{Li:2010ts,Wang:2013cia,He:2015pra}.  
%The effective strong coupling constant $\alpha_s=g^{2}/4\pi$ in LBT is fixed and fitted to experimental data.

The CLVisc hydrodynamic model \cite{Pang:2012he,Pang:2018zzo} is used to provide the space-time profile of the local temperature and fluid velocity in the QGP medium. In the linear approximation ($\delta f\ll f$), interaction among jet shower and recoil partons is neglected and only interaction with medium partons is allowed. The LBT model has been able to describe experimental data on suppression of single inclusive light and heavy quark hadrons \cite{Cao:2016gvr,Cao:2017hhk}, $\gamma$-hadron correlations \cite{Chen:2017zte}, single inclusive jets \cite{He:2018xjv}, and $\gamma/Z^0$-jet correlations \cite{Wang:2013cia,Luo:2018pto,Zhang:2018urd}.

\noindent{\it 5. Gradient tomography of $\gamma$-jets}: --
In this study we use LBT simulations of $\gamma$-jet events in 0-10\% central Pb+Pb collisions at $\sqrt{s}=2.76$ TeV to illustrate the gradient tomography of jet quenching. We use PYTHIA 8~\cite{Sjostrand:2006za,Sjostrand:2007gs} to generate the initial jet shower partons in $\gamma$-jet events in $p+p$ collisions and let these jet shower partons propagate through the QGP medium according to the LBT model. Events are generated first with a minimum transverse momentum transfer of the hard processes $\hat p_T>30$ GeV/$c$.  Only those events that have $p_T^\gamma=60-80$ GeV/$c$ are selected for the final analysis. 

The event-by-event initial conditions from A multiphase transport (AMPT) model \cite{Lin:2004en} are used for the evolution of the background QGP medium according to the CLVisc (3+1)D viscous hydrodynamics model ~\cite{Pang:2012he,Pang:2018zzo}.  The initial energy-momentum density is normalized at the initial time $\tau_0=0.2$ fm/$c$ so that the final bulk hadron spectra with freeze-out temperature $T_{\rm f}=137$ MeV can reproduce experimental data on the charged hadron rapidity and transverse momentum distributions \cite{Pang:2018zzo}. For more details of LBT simulations of $\gamma$-jet events in heavy-ion collisions we refer readers to Ref.~\cite{Luo:2018pto}. The FastJet \cite{Cacciari:2011ma} is used to reconstruct jets from the final partons within LBT.  

For each hydro event, one can determine the second-order event plane  angle,
\begin{equation}
\Psi_2=\frac{1}{2}\arctan\frac{\langle p_T\sin(2\phi)\rangle}{\langle p_T\cos(2\phi)\rangle},
\end{equation}
which is a good approximation of the reaction plane \cite{Pang:2012he}. One can select the trigger photons to have a given direction relative to this event plane. In this study, we let the trigger photon align with the second-order event plane (unless specified). One can then calculate the transverse momentum asymmetry $A_{E_\perp}^{\vec n}$, according to Eq.~(\ref{eq:asym}) but for parton's transverse energy instead of number distribution, for each $\gamma$-jet event. To focus on jet shower partons, we impose a cut $p_T>3$ GeV/$c$ on the final partons from the LBT simulation. The final results are averaged over 200 hydro events and $10^5$  $\gamma$-jet events with $p_T^\gamma=60-80$ GeV/$c$ with each hydro event.

\begin{figure}
\includegraphics[width=4.6cm]{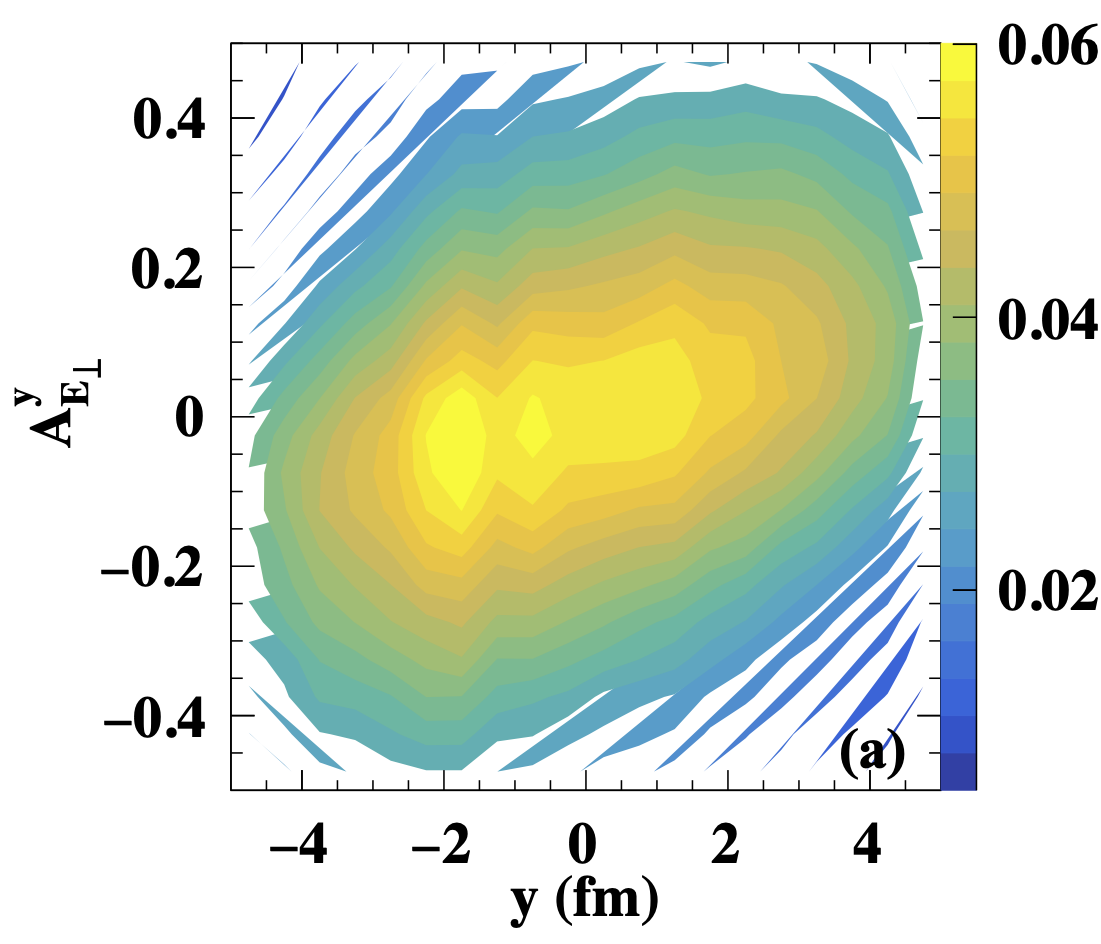}
\includegraphics[width=3.9cm]{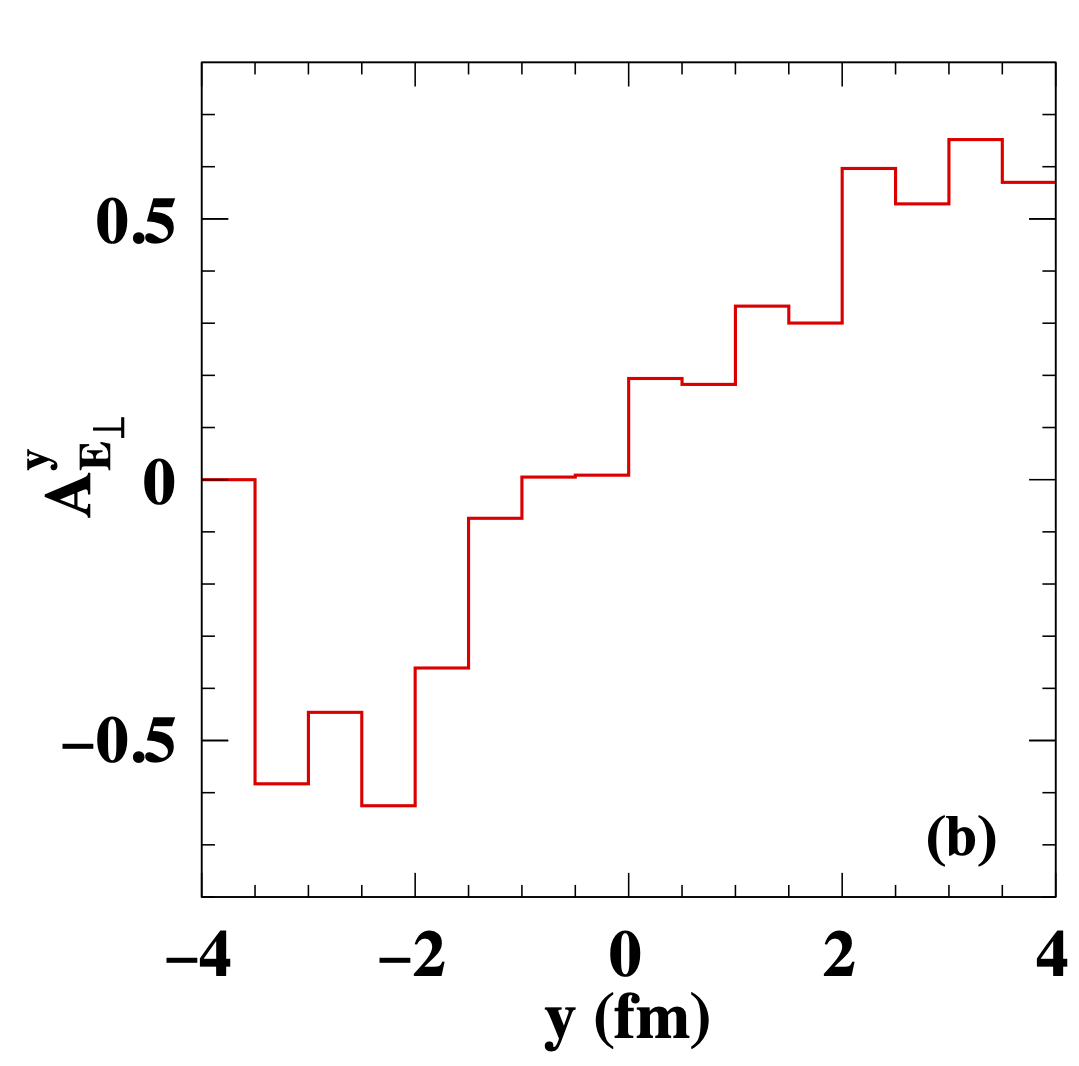}
\vspace{-0.2 in}
 \caption{(a) (Color online) LBT event distribution in the transverse asymmetry $A_{E_\perp}^y$ and the transverse coordinate $y$ and (b) mean value of $A_{E_\perp}^y$ as a function of $y$ for $\gamma$-triggered jets with $p_T^\gamma=60-80$ GeV/$c$ and $p_T^{\rm jet}=80-90$ GeV/$c$ in 0-10\% central Pb+Pb collisions at $\sqrt{s}=2.76$ TeV. }
 \label{asym-y}
\end{figure}

Shown in Fig.~\ref{asym-y}(a) is the event number distribution in the transverse momentum asymmetry $A_{E_\perp}^y$ and the initial transverse coordinate $y$ of $\gamma$-jet production in 0-10\% central Pb+Pb collisions at $\sqrt{s}=2.76$ TeV.  At a given transverse position $y$ relative to the center of the overlapped nuclei, the transverse asymmetry has a shifted distribution due to the drift in the transverse momentum of jet shower partons caused by the transverse gradient of the jet transport coefficient $\hat q$.  Shown in Fig.~\ref{asym-y}(b) is the mean value of $A_{E_\perp}^y$  as a function of $y$. One can clearly see an approximately linear correlation between the transverse asymmetry $A_{E_\perp}^y$  and the initial transverse coordinate of the jet production.

\begin{figure}
\includegraphics[width=8.0cm]{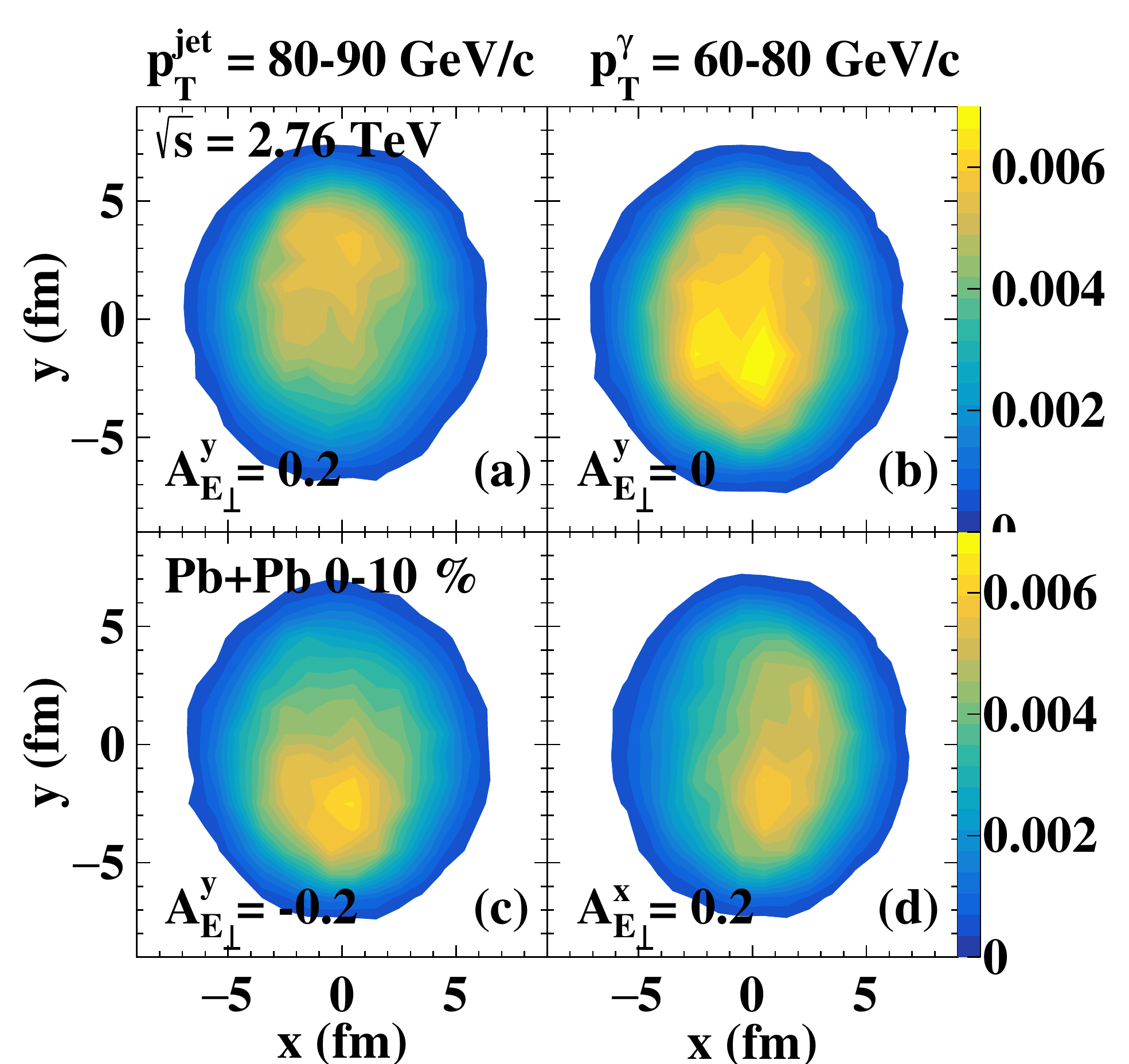}
\vspace{-0.0 in}
 \caption{(Color online) LBT transverse distribution of the initial $\gamma$-jet production position with a given value of transverse asymmetry  (a) $A_{E_\perp}^y=0.2$ (b) $A_{E_\perp}^y=0$ (c) $A_{E_\perp}^y=-0.2$, and (d) $A_{E_\perp}^x=0.2$ when the trigger photon is perpendicular to the event plane with $p_T^\gamma=60-80$ GeV/$c$ and $p_T^{\rm jet}=80-90$ GeV/$c$ in 0-10\% central Pb+Pb collisions at $\sqrt{s}=2.76$ TeV. }
 \label{x-y-asym}
\end{figure}

For a given range of the transverse asymmetry $A_{E_\perp}^y$, we can also calculate the contour distribution of the initial $\gamma$-jet position in the transverse plane as shown in Figs.~\ref{x-y-asym}(a)--(c). For positive (negative) values of $A_{E_\perp}^y$, the initial $\gamma$-jet production position is distributed around a region above (below) and away from the center of the overlapped nuclei. By selecting the value of the transverse asymmetry, one can therefore localize the transverse position of the initial jet production position for a more detailed study of jet quenching and the properties of the QGP matter. Such a procedure can be repeated for different orientations of the trigger particle (photon in the $\gamma$-jet case) relative to the event plane, as shown in Fig.~\ref{x-y-asym}(d) when the trigger photon is perpendicular to the event plane. This is essentially a systematic scanning of the whole volume of the QGP matter with the gradient tomography and related jet quenching measurements.

\noindent{\it 6. Summary and Discussions}: --
Based on the diffusion and drift of jet partons propagating in a nonuniform QGP medium and the resultant momentum asymmetry, we have proposed a gradient tomography of jet quenching.  The transverse asymmetry along the direction of the gradient of the jet transport coefficient is shown to correlate almost linearly with the initial  transverse position of the propagating parton. Using the LBT model for event-by-event simulations of $\gamma$-triggered jets in Pb+Pb collisions at $\sqrt{s}=2.76$ TeV, we further demonstrate that one can indeed use the transverse asymmetry to localize the initial jet production position for detailed study of jet quenching with a given jet propagation path. Varying the orientation of the trigger particle relative to the event plane, one can effectively scan the whole volume of the QGP with jet quenching study. Many measurements with this gradient tomography are possible. For example, one can define the net transverse momentum shift
\begin{equation}
\Delta k_\perp^{\vec n}=\frac{\int d^3rd^3kf_a(\vec k,\vec r) \vec k\cdot \vec n}{\int d^3rd^3kf_a(\vec k,\vec r)},
\label{eq:asymk}
\end{equation}
which should be proportional to the average value of $\hat q$ along the propagation path. It therefore can be used to measure $\hat q$ for a given propagation path as specified by the transverse momentum asymmetry $A_N^{\vec n}$.  
This method can also be applied to jet production in electron ion collider (EIC) for a more detailed study of gluon saturation and jet transport in cold nuclei. As an added bonus, the simple but nontrivial analytic solution of the diffusion and drift equation in a uniform medium we discovered in this study will be useful for many transport and hybrid hydro-transport problems where the diffusion of a particle within a short period of relaxation time needs to be incorporated.

We thank T. Luo and W. Ke for discussions, and W. Chen and Z. Yang for providing CLVisc hydro events. This work is supported by NSFC under Grants No. 11935007, No. 11221504,  and No. 11890714, by DOE under Contract No. DE-AC02-05CH11231, and by NSF under Grant No. ACI-1550228 within the JETSCAPE Collaboration. Computations are performed at DOE NERSC and NSC$^3$ of CCNU..

  \end{document}